\documentclass{elsart}
\usepackage{graphicx}
\usepackage{amsmath}
\usepackage{amssymb}
\newcommand{\ie}{i.e.}
\newcommand{\nn}{n.n.}
\newcommand{\nnn}{n.n.n.}
\begin{document}
\runauthor{Derzhko and J{\c{e}}drzejewski}
\begin{frontmatter}

\title{From phase separation to long-range order in a system of
interacting electrons}

\author[Wroclaw]{Volodymyr Derzhko} and
\author[Wroclaw,Lodz]{Janusz J{\c{e}}drzejewski\thanksref{email}}

\address[Wroclaw]{Institute of Theoretical Physics,
University of Wroc{\l}aw, pl. Maksa Borna 9, 50--204 Wroc{\l}aw,
Poland}
\address[Lodz]{Department of Theoretical Physics, University of
{\L}{\'{o}}d{\'{z}}, ul. Pomorska 149/153, 90--236
{\L}\'{o}d\'{z}, Poland}

\thanks[email]{Corresponding author: J. J{\c{e}}drzejewski, phone:
+48 71 3759415, fax: +48 71 3214454, e-mail: jjed@ift.uni.wroc.pl}

\begin{abstract}
We study a system composed of fermions (electrons), hopping on a
square lattice, and of immobile particles (ions), that is
described by the spinless Falicov--Kimball Hamiltonian augmented
by a next-nearest-neighbor attractive interaction between the ions
(a nearest-neighbor repulsive interaction between the ions can be
included and does not alter the results). A part of the
grand-canonical phase diagram of this system is constructed
rigorously, when the coupling between the electrons and ions is
much stronger than the hopping intensity of electrons. The
obtained diagram implies that, at least for a few rational
densities of particles, by increasing the hopping intensity the
system can be driven from a state of phase separation to a state
with a long-range order. This kind of transitions occurs also,
when the hopping fermions are replaced by hopping hard-core
bosons.
\end{abstract}

\end{frontmatter}

\section{Introduction}
Since forty years, it is the Hubbard or the extended Hubbard
models that are most frequently studied when properties of
strongly correlated electrons are to be investigated. The central
issue is the phase diagram of these models. Despite the apparent
simplicity of both models and concerted efforts of many
researchers, the complicated structure of these phase diagrams,
even at zero temperature, has not been revealed completely and
unquestionably; rigorous results are scarce. A few years ago
Nakamura \cite{Nakamura} predicted a new kind of quantum phase
transition in the ground state of a half-filled extended Hubbard
model chain. On a line, in the space of the two relevant
interaction parameters (representing on-site and nearest-neighbor
({\nn}) repulsion due to Coulomb forces, expressed in the units of
the hopping intensity), for large values of these parameters, the
system is in the phase-separated state, which is a mixture of a
charge-density-wave phase and a spin-density-wave phase. The
transition occurs when the parameters are decreased along the
line, at a critical point where the parameters assume intermediate
values (of the order of one). Thus, the transition region is
hardly accessible by perturbation methods. A new state exhibits a
long-range order, the so called bond-order-wave. There are many
aspects of the phase diagram to be studied in connection with the
new conjectured transition, and they are vigorously discussed in
the physics literature recently \cite{SSC1,Jeckel,SSC2}. In this
paper we address just one aspect of the new transition that we
find remarkable. Namely, by increasing the hopping intensity of
electrons the system is driven from a phase-separated state to a
crystalline state with a long-range order. Our aim is to construct
a model of interacting electrons (by the way we consider an
analogous model of interacting hard-core bosons), where the
existence of a transition exhibiting this feature can be
demonstrated rigorously. The model to be studied is a simplified
version of the one band, spin $1/2$ Hubbard model, known as the
static approximation (one sort of electrons hops while the other
sort is immobile), with the Hamiltonian $H_{FK}$, extended by a
next-nearest-neighbor ({\nnn}) attractive interaction between the
immobile particles, given by the Hamiltonian $V$. Thus the total
Hamiltonian of the system reads:

\begin{eqnarray}
H_0  =  H_{FK} + V, \label{H0} \\
H_{FK}  =  -t \sum\limits_{\langle x,y \rangle_{1}} \, \left(
c^{+}_{x}c_{y}+h.c. \right) + U\sum\limits_{x}\left(
c^{+}_{x}c_{x} -
\frac{1}{2} \right) s_{x} ,  \label{HFK} \\
V  =  -\frac{{\tilde{\varepsilon}}}{16} \sum\limits_{\langle x,y
\rangle_{2}} s_{x}s_{y}, \ \ \ t>0, \ \tilde{\varepsilon} > 0.
\label{V}
\end{eqnarray}

In the above formulae, the underlying lattice is a square
lattice, denoted $\Lambda$, consisting of sites $x, y, \ldots$
whose number is $|\Lambda|$, having the shape of a
$\sqrt{|\Lambda|} \times \sqrt{|\Lambda|}$ torus. In
(\ref{HFK},\ref{V}) and below, the sums $\sum_{\langle x,y
\rangle_{i}}$, $i=1,2,3$, stand for the summation over all the
$i$-th \nn{} pairs of lattice sites, with each pair counted once.

The subsystem of mobile spinless electrons is described in  terms
of creation and annihilation operators of an electron at site
$x$: $c^{+}_{x}$, $c_{x}$, respectively, satisfying the canonical
anticommutation relations, with $t$ being the \nn{} hopping
intensity. The total electron-number operator is $N_{e} =
\sum_{x} c^{+}_{x} c_{x}$, and (with a little abuse of notation)
the corresponding electron density is $\rho_{e} =
N_{e}/|\Lambda|$.

The subsystem of ions is described by a collection of pseudo-spins
$\left\{ s_{x} \right\}_{x \in \Lambda}$, with $s_{x} = 1, -1$
($1$ if the site $x$ is occupied by an ion and $-1$ if it is
empty), called the {\em ion configurations}. The total number of
ions is $N_{i} = \sum_{x} ( s_{x} + 1 )/2$ and the ion density is
$\rho_{i} = N_{i}/|\Lambda|$. In contradistinction to the
electron subsystem, the ions interact directly: two ions that
occupy two \nnn{} sites attract each other, contributing the
energy $-\tilde{\varepsilon}/16$, with $\tilde{\varepsilon} > 0$.

Clearly, in the composite system, whose Hamiltonian is given by
(\ref{H0}), with arbitrary electron-ion coupling $U$, the
particle-number operators $N_{e}$, $N_{i}$, and pseudo-spins
$s_{x}$, are conserved. Therefore the description of the classical
subsystem in terms of the ion configurations $S =\left\{ s_{x}
\right\}_{x \in \Lambda}$ remains valid. Whenever periodic
configurations of pseudo-spins are considered it is assumed that
$\Lambda$ is sufficiently large, so that it accommodates an
integer number of elementary cells.

Nowadays, $H_{FK}$ is widely known as the Hamiltonian of the
spinless Falicov--Kimball model, a simplified version of the
Hamiltonian put forward in \cite{FK}. A lot of results, including
rigorous ones, like a proof of the existence of a phase transition
\cite{BS,KL} for instance, have been obtained for the system
described by this Hamiltonian (a review and an extensive list of
references can be found in \cite{GM,JL}).

In what follows, we shall study the ground-state phase diagram of
the system defined by (\ref{H0}) in the grand-canonical ensemble.
That is, let
\begin{equation}
H \left( \mu_{e}, \mu_{i} \right) = H_{0} - \mu_{e}N_{e} -
\mu_{i}N_{i}, \label{Hmu}
\end{equation}
where $\mu_{e}$, $\mu_{i}$ are the chemical potentials of the
electrons and ions, respectively, and let $E_{S}\left( \mu_{e},
\mu_{i} \right)$ be the ground-state energy of $H \left( \mu_{e},
\mu_{i} \right)$, for a given configuration $S$ of the ions.
Then, the ground-state energy of $H \left( \mu_{e}, \mu_{i}
\right)$, $E_{G}\left( \mu_{e}, \mu_{i} \right)$,  is defined as
$E_{G}\left( \mu_{e}, \mu_{i} \right) = \min \left\{ E_{S} \left(
\mu_{e}, \mu_{i} \right): S \right\}$. The minimum is attained at
the set $G$ of the ground-state configurations of ions. We shall
determine the subsets of the $\left( \mu_{e}, \mu_{i}
\right)$-plane, where $G$ consists of periodic configurations of
ions, uniformly in the size of the underlying square lattice.

The paper is organized as follows. In the first subsection of the
next section we provide a number of symmetry properties of the
ground-state energy, $E_{S} \left( \mu_{e}, \mu_{i} \right)$,
which facilitate the studies of the phase diagram that follow, and
give the strong-coupling expansion of $E_{S} \left( \mu_{e},
\mu_{i} \right)$ in the case of hopping fermions (subsection 2.1).
Then, in subsection 2.2 we construct the grand-canonical phase
diagram due to the ground-state energy truncated at fourth order
of the strong-coupling expansion in the case of hopping fermions
(this is sufficient to draw our conclusions). After that, in
subsection 2.3 we consider the grand-canonical phase diagram due
to the ground-state energy truncated at fourth order of the
strong-coupling expansion in the case of hopping hard-core bosons.
Finally, in Section~3 we discuss the implications of the phase
diagram obtained with the truncated effective interaction, when
the remainder of the strong-coupling expansion is taken into
account.

\section{Grand-canonical phase diagram in the strong coupling
regime}

\subsection{Properties of $E_{S} \left( \mu_{e}, \mu_{i} \right)$ and
its strong-coupling expansion}

In studies of grand-canonical phase diagrams an important role is
played by unitary transformations that exchange particles and
holes: $c^{+}_{x}c_{x} \rightarrow 1 -  c^{+}_{x}c_{x}$ and $s_{x}
\rightarrow -s_{x}$, and for some $\left( \mu^{0}_{e}, \mu^{0}_{i}
\right)$ leave the Hamiltonian  $H \left( \mu_{e}, \mu_{i}
\right)$ invariant. For the electron subsystem such a role is
played by the transformations: $c_{x}^{+} \rightarrow \epsilon_{x}
c_{x}$, with $\epsilon_{x} = 1$ at the even sublattice of
$\Lambda$ and $\epsilon_{x} = -1$ at the odd one. Clearly, since
$H_{0}$ is invariant under the joint hole--particle transformation
of electrons and ions, $H \left( \mu_{e}, \mu_{i} \right)$ is
hole--particle invariant at the point $(0,0)$. At the
hole--particle symmetry point, the system under consideration has
very special properties, which simplify studies of its phase
diagram \cite{KL}. Moreover, by means of the defined above
hole--particle transformations one can determine a number of
symmetries of the grand-canonical phase diagram \cite{GJL}. The
peculiarity of the model is that the case of attraction ($U<0$)
and the case of repulsion ($U>0$) are related by a unitary
transformation (the hole--particle transformation for ions): if
$S$ is a ground-state configuration at $\left( \mu_{e}, \mu_{i}
\right)$ for $U>0$, then $-S$ is the ground-state configuration at
$\left( \mu_{e}, -\mu_{i} \right)$ for $U<0$. Consequently,
without any loss of generality one can fix the sign of the
coupling constant $U$. Moreover (with the sign of $U$ fixed),
there is the {\em inversion symmetry\/} of the grand-canonical
phase diagram, that is, if $S$ is a ground-state configuration at
$\left( \mu_{e}, \mu_{i} \right)$, then $-S$ is the ground-state
configuration at $\left( -\mu_{e}, -\mu_{i} \right)$. Therefore,
it is enough to determine the phase diagram in the half-plane
specified by fixing the sign of one of the chemical potentials.
Additional properties emerge in the {\em strong-coupling regime},
\ie{} for $|U|>4t$ \cite{GJL}. Specifically, if the electron
chemical potential is in the open interval $\left| \mu_{e} \right|
< |U| - 4t$, then for any two ion configurations $S$, $S^{\prime}$
the energy difference $E_{S} \left( \mu_{e}, \mu_{i} \right) -
E_{S^{\prime}} \left( \mu_{e}, \mu_{i} \right)$ is constant along
the lines $\mu_{e} \pm \mu_{i} = {\mathrm{const}}$, with minus
sign referring to the case of positive $U$ (repulsion) while the
plus sign --- to the case of negative $U$ (attraction). This
property together with the inversion symmetry implies that in the
stripe $\left| \mu_{e} \right| < |U| - 4t$ of the $\left( \mu_{e},
\mu_{i} \right)$-plane it is enough to determine the phase
diagram at the half-line $\mu_{e}=0$, $\mu_{i} < 0$ (or $\mu_{i}
>0$).

Our aim in this paper is to investigate the ground-state phase
diagram off the symmetry point. According to the state of the art,
this is feasible only in the strong-coupling regime. From now on,
we shall consider exclusively the case of the strong positive
coupling, \ie{} $U/t >4$. In this case, it is convenient to
express all the parameters of $H \left( \mu_{e}, \mu_{i}
\right)$, in the units of $|U|$, \ie{} we change $t \to t/|U|$,
$\tilde{\varepsilon} \to \tilde{\varepsilon}/|U|$, etc, but keep
the previous notation. Then, $H_{0}$ assumes the form
\begin{equation}
H_{0} =-t \sum\limits_{\langle x,y \rangle_{1}} \, \left(
c^{+}_{x}c_{y} + h.c. \right) + 1 \sum\limits_{x} \left(
c^{+}_{x}c_{x} - \frac{1}{2} \right) s_{x} -
\frac{{\tilde{\varepsilon}}}{16} \sum\limits_{\langle x,y
\rangle_{2}} s_{x} s_{y}, \label{H0t/U}
\end{equation}
and the strong-coupling regime is given by the condition $t <
1/4$. The possibility of studying the phase diagram in the
strong-coupling regime stems from the fact that in the open stripe
$\left| \mu_{e} \right| < 1 - 4t$, the ion density $\rho_{i}$ of a
ground-state configuration determines the electron density:
$\rho_{e} = \rho_{i}$ for $U<0$, and $\rho_{e} = 1 - \rho_{i}$ for
$U>0$. Then, one can formally expand the ground-state energy
$E_{S} \left( \mu_{e}, \mu_{i} \right)$ in powers of $t$, what
results in the so called effective Hamiltonian:
\begin{eqnarray}
\label{expansion} E_{S} \left( \mu_{e}, \mu_{i} \right) &=&
-\frac{1}{2} \left( \mu_{i}-\mu_{e} \right) \sum\limits_{x} s_{x}
-\frac{1}{2} \left( \mu_{i}+\mu_{e}+1 \right) |\Lambda| +
\nonumber \\
&&\left[ \frac{t^{2}}{4}- \frac{9t^{4}}{16} \right]
\sum\limits_{\langle x,y \rangle_{1}}  s_{x}s_{y} + \left[
\frac{3t^{4}}{16}- \frac{{\tilde{\varepsilon}}}{16} \right]
\sum\limits_{\langle x,y \rangle_{2}}  s_{x}s_{y} +
\nonumber \\
&&\frac{t^{4}}{8} \sum\limits_{\langle x,y \rangle_{3}} s_{x}s_{y}
+ \frac{t^{4}}{16} \sum\limits_{P} \left(1+5s_{P}\right) +
R^{(4)},
\end{eqnarray}
up to a term independent of the ion configuration and the chemical
potentials. In (\ref{expansion}), $P$ denotes the $(2 \times
2)$-plaquette of the square lattice $\Lambda$, $s_{P}$ stands for
the product of pseudo-spins assigned to the corners of $P$, and
the remainder $R^{(4)}$, which is independent of the chemical
potentials and ${\tilde{\varepsilon}}$, collects those terms of
the expansion that are proportional to $t^{2m}$, with
$m=3,4,\ldots$. The expansion (\ref{expansion}) (with
$\tilde{\varepsilon}=0$) was introduced and the phase diagram,
according to the expansion truncated at the fourth order, was
studied in \cite{GJL}. It turns out however that, in the
strong-coupling regime the expansion (\ref{expansion}) is
absolutely convergent, uniformly in $\Lambda$
\cite{Kennedy1,GMMU}. Due to this fact, it is possible to
establish rigorously a part of the phase diagram (that is the
ground states are determined everywhere in the $\left( \mu_{e},
\mu_{i} \right)$-plane, except some regions), by determining the
phase diagram of the expansion truncated at the order $k$, that is
according to the effective Hamiltonian $E^{(k)}_{S} \left(
\mu_{e}, \mu_{i} \right)$.

\subsection{Construction of the phase diagram up to fourth order.
The case of hopping fermions}

\begin{figure}
\includegraphics[width=0.47\textwidth]{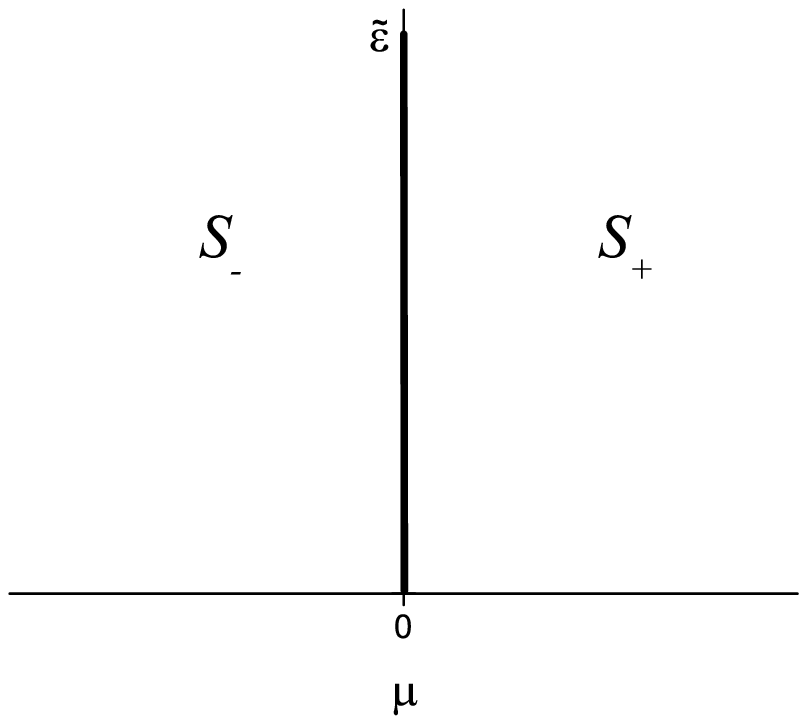}
\hfill
\includegraphics[width=0.47\textwidth]{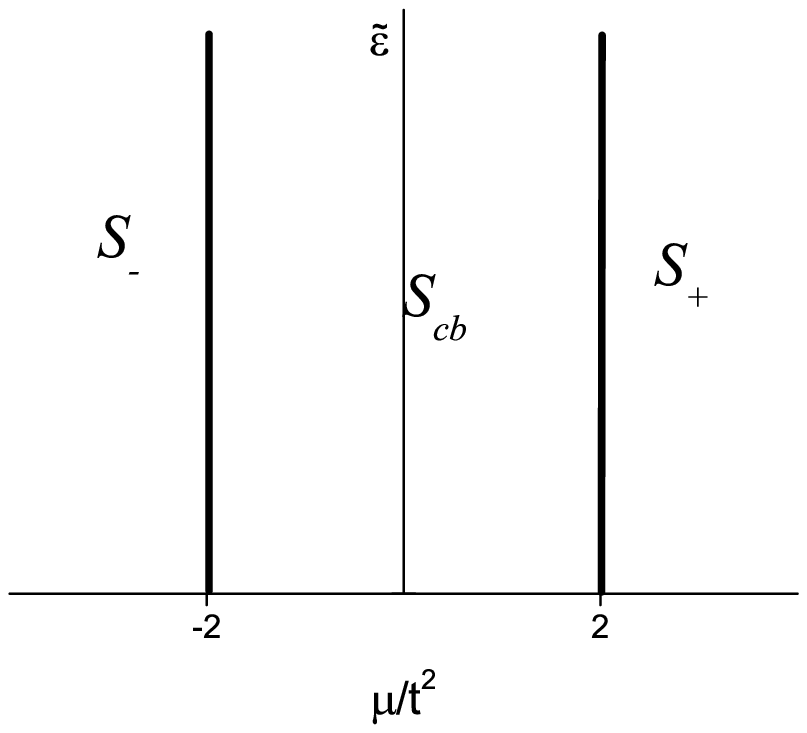}
\\
\parbox[t]{0.47\textwidth}{\caption{Ground-state
phase diagram of $E^{(0)}_{S} \left( 0, \mu \right)$.} \label{t0}}
\hfill
\parbox[t]{0.47\textwidth}{\caption{Ground-state
phase diagram of  $E^{(2)}_{S} \left( 0, \mu \right)$.}
\label{t2}}
\end{figure}

To construct the phase diagram according to the effective
Hamiltonian $E^{(k)}_{S} \left( \mu_{e}, \mu_{i} \right)$, we use
the $m$-potential method introduced in \cite{Slawny}, with
technical developments given in \cite{GJL,Kennedy1,GMMU}. By
virtue of the comments in the preceding subsection, in what
follows we set $\mu_{e}=0$ and $\mu_{i} = \mu$. In the order zero
the effective Hamiltonian reads:
\begin{eqnarray}
\label{E0} E^{(0)}_{S} \left( 0, \mu \right) &=&-\frac{\mu}{2}
\sum\limits_{x} \left( s_{x}+1 \right) -
\frac{{\tilde{\varepsilon}}}{16}
\sum\limits_{\langle x,y \rangle_{2}} s_{x}s_{y}\\
&=& \sum\limits_{P} H_{P}^{(0)} \nonumber,
\end{eqnarray}
where
\begin{equation}
\label{HP0} H_{P}^{(0)}= -\frac{\mu}{8} \sum\limits_{x} \left(
s_{x}+1 \right) - \frac{{\tilde{\varepsilon}}}{16}
\sum\limits_{\langle x,y \rangle_{2}} s_{x}s_{y}.
\end{equation}
Except the point $\mu=0$, $\tilde{\varepsilon}=0$, where all the
configurations have the same energy, the plaquette potentials
$H_{P}^{(0)}$ are minimized by the restrictions to $P$ of a few
periodic configurations on $\Lambda$. For $\mu<0$ ($\mu>0$) it is
the {\em ferromagnetic  configuration\/} $S_{-}$ ($S_{+}$), where
$s_{x}=-1$ at every site ($s_{x}=+1$ at every site), while at the
half-line $\mu=0$, $\tilde{\varepsilon}>0$, besides the
ferromagnetic configurations $S_{-}$ and $S_{+}$, these are the
two {\em antiferromagnetic (or chessboard) configurations\/}
$S_{cb}^{e}$, where $s_{x}=\epsilon_{x}$, and $S_{cb}^{o}$, where
$s_{x}=-\epsilon_{x}$, with $\epsilon_{x}=1$ if $x$ belongs to the
even sublattice of $\Lambda$ and $\epsilon_{x}=-1$ otherwise.
Moreover, out of the restrictions $S_{-|P}$, $S_{+|P}$,
$S^{e}_{cb|P}$, and $S_{cb|P}^{o}$ only four ground-state
configurations can be built, which coincide with the four
configurations named above. Clearly, this is due to the \nnn{}
ferromagnetic interaction. The phase diagram according to the
effective Hamiltonian $E^{(0)}_{S} \left( 0, \mu \right)$ is shown
in Fig.~\ref{t0}.

In the next order, which takes into the account interactions
proportional to $t^{2}$, a \nn{} antiferromagnetic interaction
appears. The effective Hamiltonian reads
\begin{eqnarray}
\label{E2} E^{(2)}_{S} \left( 0, \mu \right) &=&-\frac{\mu}{2}
\sum\limits_{x} \left( s_{x}+1 \right) + \frac{t^{2}}{4}
\sum\limits_{\langle x,y \rangle_{1}} s_{x}s_{y} -
\frac{{\tilde{\varepsilon}}}{16} \sum\limits_{\langle x,y
\rangle_{2}}
s_{x}s_{y} \\
&=& \sum\limits_{P} H_{P}^{(2)}, \nonumber
\end{eqnarray}
where,
\begin{equation}
\label{HP2} H_{P}^{(2)}= -\frac{\mu}{8} \sum\limits_{x} \left(
s_{x}+1 \right) + \frac{t^{2}}{8} \sum\limits_{\langle x,y
\rangle_{1}} s_{x}s_{y} - \frac{{\tilde{\varepsilon}}}{16}
\sum\limits_{\langle x,y \rangle_{2}} s_{x}s_{y}.
\end{equation}
By determining the plaquette configurations that minimize the
potentials $H_{P}^{(2)}$ one finds the ground-state phase diagram
shown in Fig.~\ref{t2}. The boundary between $S_{+}$ and $S_{-}$,
determined in the order zero, \ie{} the half-line $\mu=0$,
$\tilde{\varepsilon}>0$, is replaced by the stripe of the width
$4t^{2}$, centered at this half-line. Outside this stripe the
ground state remains ferromagnetic. Inside the stripe, including
its boundary on the line ${\tilde{\varepsilon}}=0$, there are
only two ground-state configurations, the antiferromagnetic ones.
At the boundary of the stripe, given by $\mu=-2t^{2},
{\tilde{\varepsilon}}>0$ ($\mu=2t^{2}, {\tilde{\varepsilon}}>0$),
there are exactly three ground-state configurations $S_{-}$,
$S_{cb}^{e}$, and $S_{cb}^{o}$ ($S_{+}$, $S_{cb}^{e}$, and
$S_{cb}^{o}$). Only at the points $(\mu=\pm 2t^{2},
{\tilde{\varepsilon}}=0)$ the number of ground-state
configurations grows exponentially with the size of the
underlying lattice (like $\exp({\mathrm{const}}|\Lambda|)$). The
Hamiltonian $E^{(2)}_{S}\left( 0, \mu \right)$ is known in the
physics literature as the Fisher-stabilized Ising antiferromagnet
\cite{FILS}.

To proceed further and analyze the effect of the fourth-order
interactions, we make use of the inversion symmetry that enables
us to fix, without any loss of generality, the sign of the
chemical potential. From now on, we restrict our investigations to
the case $\mu <0$.

Let us emphasize that we are interested in the phase diagram for
sufficiently small $t$. Therefore clearly, it is the neighborhood
of radius $O(t^{4})$ of the point $\mu=\pm 2t^{2},
{\tilde{\varepsilon}}=0$, in the
$(\mu,{\tilde{\varepsilon}})$-plane, where the effect of the
fourth order interactions can be most significant. In this
neighborhood, it is convenient to introduce the new coordinates,
$\delta$, $\varepsilon$, as follows:
\begin{equation}
\label{d,e} \mu=-2t^{2} + \delta t^{4}, \ \ \
{\tilde{\varepsilon}}=\varepsilon t^{4}.
\end{equation}
In terms of $\delta$, $\varepsilon$, the effective Hamiltonian up
to the order four can be written in the form:
\begin{eqnarray}
\label{E4} E^{(4)}_{S} \left( 0,\delta \right) & = &
\frac{t^{2}}{4} \sum\limits_{\langle x,y \rangle_1} \left(
s_{x}s_{y}+s_{x}+s_{y}+1 \right) + \frac{t^{4}}{2} \sum\limits_{T}
H^{(4)}_{T} =\\
& & t^{2} \sum\limits_{\substack{\langle x,y \rangle_{1} \\
s_{x}=s_{y}=1}} 1 + \frac{t^{4}}{2} H^{(4)}_{{\mathrm{eff}}},
\nonumber
\end{eqnarray}
with
\begin{eqnarray}
\label{HT} H^{(4)}_{T} &=& -\delta  \left( s_{5} + 1 \right)
-\frac{3}{16}\sum\limits_{\langle x,y \rangle_1} s_{x}s_{y} +
\left[ \frac{3}{32}-\frac{\varepsilon}{32} \right]
\sum\limits_{\langle x,y \rangle_{2}} s_{x}s_{y} + \\
& & \frac{1}{12} \sum\limits_{\langle x,y \rangle_{3}} s_{x}s_{y}
+ \frac{1}{32} \sum\limits_{P} \left( 5s_{P}+1 \right) , \nonumber
\end{eqnarray}
where we have omitted the term $t^{2}|\Lambda|/2$. In
(\ref{E4},\ref{HT}), $T$ stands for a $(3 \times 3)$-plaquette
(later on called the $T$-plaquette) of the square lattice, whose
sites are labeled from the left to the right, starting at the
bottom left corner, so that $s_{5}$ is the pseudo-spin of the
central site and its left-neighbor pseudo-spin is $s_{4}$. The
above form of $E^{(4)}_{S}\left( 0,\delta \right)$ shows
manifestly that, in a neighborhood of the point $\delta=0$,
$\varepsilon=0$ whose radius is $O(1)$, those configurations that
contain pairs of \nn{} sites, $\langle x,y \rangle_{1}$, with
pseudo-spins $s_{x}=s_{y}=1$ cannot be the ground state
configurations (their energy is larger than the energy of the
configurations that do not contain such pairs by a large energy
of the order $O(t^{2})$). Therefore, in the considered region of
the phase diagram, all the effects due to the interactions up to
fourth order are described by the effective Hamiltonian
$H^{(4)}_{{\mathrm{eff}}}$, and the only admissible
configurations are those that do not contain any pairs of \nn{}
sites with pseudo-spins taking the value $1$.

\begin{figure}
\centering \includegraphics[width=\textwidth]{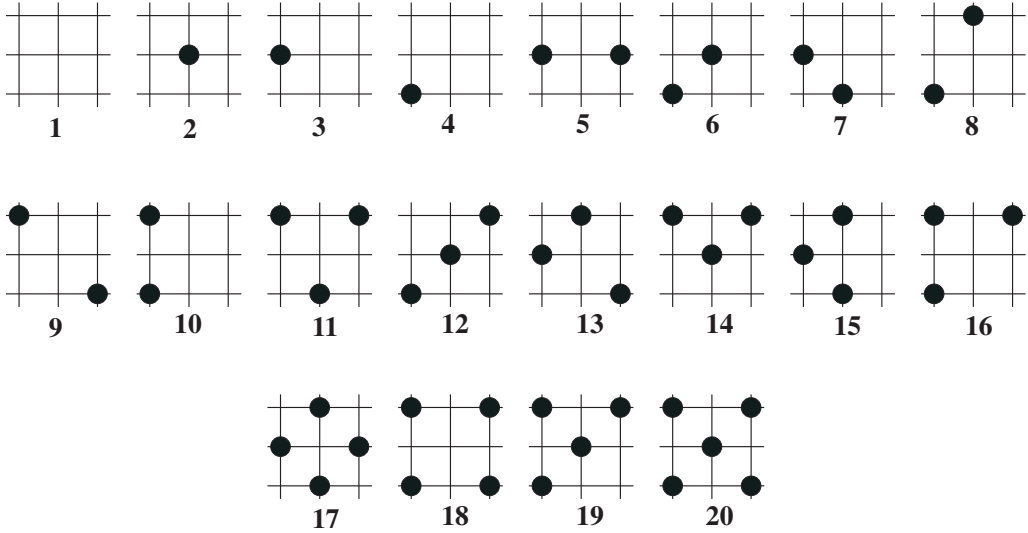}
\caption{All the admissible $T$-plaquette configurations (up to
symmetries).} \label{blocks}
\end{figure}

\begin{figure}
\centering \includegraphics[width=\textwidth]{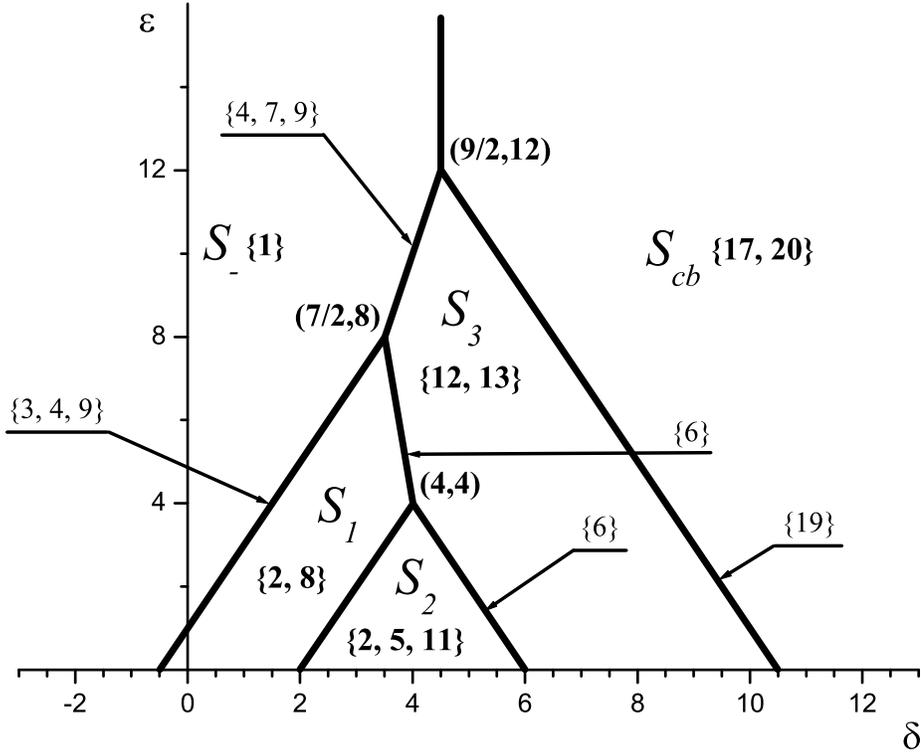}
\caption{Ground-state phase diagram of
$H^{(4)}_{{\mathrm{eff}}}$. By the crossing points of
boundary-line segments we give their coordinates. The numbers in
curly brackets, displayed by the arrows pointing towards the
boundary-line segments, identify the additional minimizing
$T$-plaquette configurations. For more comments see the text.}
\label{t4}
\end{figure}

\begin{figure}
\centering \includegraphics[width=\textwidth]{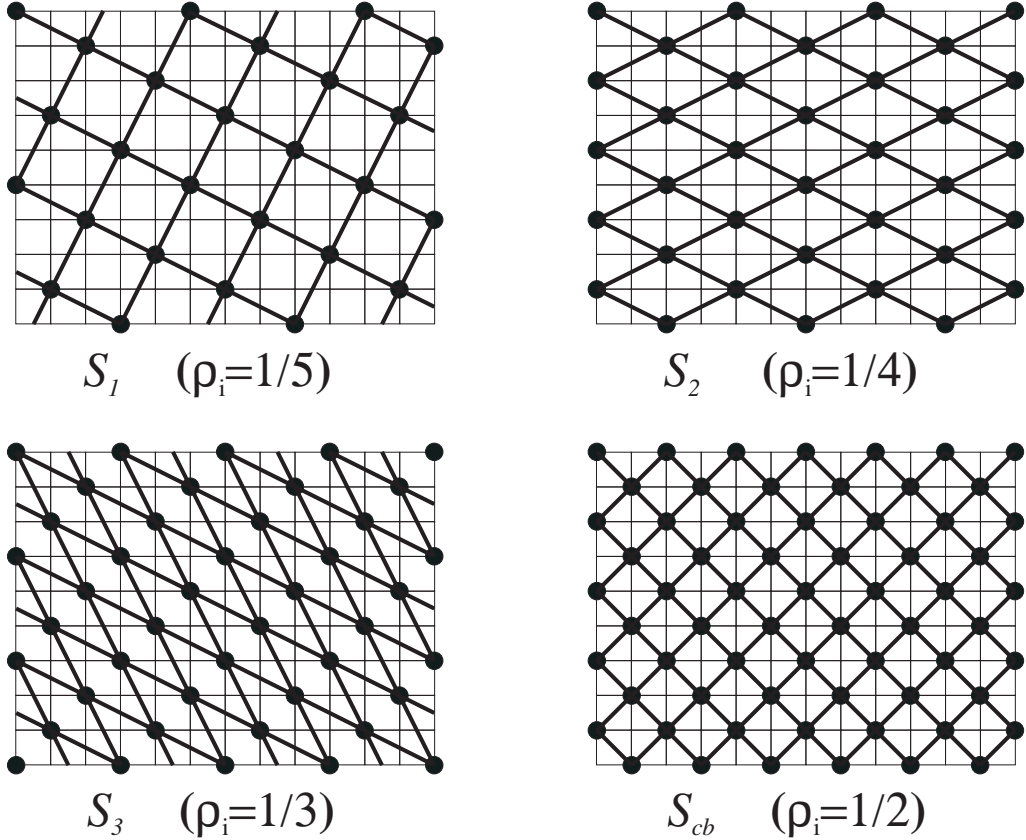}
\caption{Ground-state configurations, which appear for $\mu<0$.}
\label{configurations}
\end{figure}

To determine the phase diagram due to $H^{(4)}_{{\mathrm{eff}}}$
we use, as before, the $m$-potential method. However, in contrast
to the lower-order cases the analysis is not that
straightforward. The potentials $H^{(4)}_{T}$ cannot serve as the
$m$-potentials in the whole $\left( \delta,\varepsilon
\right)$-plane. This difficulty can be overcome by introducing
the so called zero-potentials \cite{GJL,Kennedy1,GMMU}. Here, we
follow closely \cite{Kennedy1,GMMU} and introduce the
zero-potentials $K^{(4)}_{T}$ such that they are invariant with
respect to the symmetries of $H_{0}$ and satisfy the condition
\begin{equation}
\label{K4T} \sum\limits_{T} K^{(4)}_{T} = 0.
\end{equation}
Moreover, the zero-potentials $K^{(4)}_{T}$ are chosen in the
following form
\begin{equation}
K^{(4)}_{T}=\sum\limits_{i=1}^{5} \alpha_{i}k_{T}^{(i)},
\end{equation}
where $\alpha_{i}$ are real linear functions of $\delta$ and
$\varepsilon$ (to be determined in the process of constructing
the phase diagram), while the potentials $k_{T}^{(i)}$ are given
by
\begin{eqnarray}
k_{T}^{(1)} & = & s_{1}+s_{3}+s_{7}+s_{9}-4s_{5}, \nonumber \\
k_{T}^{(2)} & = & s_{2}+s_{4}+s_{6}+s_{8}-4s_{5}, \nonumber \\
k_{T}^{(3)} & = & s_{1}s_{2}+s_{2}s_{3}+s_{3}s_{6}
+s_{6}s_{9}+s_{8}s_{9}+s_{7}s_{8}+s_{4}s_{7}+s_{1}s_{4} \nonumber\\
&& -2s_{2}s_{5}-2s_{5}s_{6}-2s_{5}s_{8}-2s_{4}s_{5}, \nonumber \\
k_{T}^{(4)} & = & s_{1}s_{5}+s_{3}s_{5}+s_{5}s_{9}
+s_{5}s_{7}-s_{2}s_{4}-s_{4}s_{8}-s_{8}s_{6}-s_{2}s_{6}, \nonumber \\
k_{T}^{(5)} & = & s_{1}s_{3}+s_{3}s_{9}+s_{7}s_{9}+s_{1}s_{7}
-2s_{4}s_{6}-2s_{2}s_{8}.
\end{eqnarray}
We note that the potentials $k_{T}^{(i)}$ are invariant with
respect to the symmetries of the Hamiltonian $H_{0}$ and they
satisfy the condition (\ref{K4T}). Consequently, the same
properties are shared by the potentials $K^{(4)}_{T}$. Now, we
can rewrite $H^{(4)}_{{\mathrm{eff}}}$ as follows
\begin{equation}
H^{(4)}_{{\mathrm{eff}}} = \sum\limits_{T} ( H^{(4)}_{T} +
K^{(4)}_{T} ),
\end{equation}
and the problem is to find, at each point $\left(
\delta,\varepsilon \right)$, the $T$-plaquette configurations
that minimize the potential $H^{(4)}_{T} + K^{(4)}_{T}$. Taking
into account the symmetries of the Hamiltonian, there are $20$
{\em admissible $T$-plaquette configurations\/} whose energies
have to be compared. These configurations are displayed and
labelled (as in \cite{Kennedy1}) in Fig.~\ref{blocks}.
The explicit analysis shows that the $\left( \delta,\epsilon
\right)$-plane decomposes into five open domains
${\mathcal{S}}_{-}$, ${\mathcal{S}}_{cb}$, ${\mathcal{S}}_{1}$,
${\mathcal{S}}_{2}$, and ${\mathcal{S}}_{3}$, whose boundaries
consist of straight-line segments, see Fig.~\ref{t4}. We shall
adopt the same symbols to denote also the sets of ground-state
configurations in these domains.
In each point $(\delta,\varepsilon)$ of a domain, it is possible
to choose the values of the coefficients $\alpha_{i}$ such that
the set of $T$-plaquette configurations minimizing the potential
$H^{(4)}_{T} + K^{(4)}_{T}$, denoted ${\mathcal{S}}_{cb|T}$,
etc., is the same. Remarkably, out of these minimizing
$T$-plaquette configurations (in a domain) one can construct
exactly one, up to the symmetries of $H_{0}$, ground-state
configuration on the lattice $\Lambda$. The obtained
configurations turn out to be periodic. Specifically, one finds
that the minimizing $T$-plaquette configurations are:
${\mathcal{S}}_{-|T}=\{1\}$, ${\mathcal{S}}_{cb|T}=\{17,20\}$,
${\mathcal{S}}_{1|T}=\{2,8\}$, ${\mathcal{S}}_{2|T}=\{2,5,11\}$,
and ${\mathcal{S}}_{3|T}=\{12,13\}$, where the numbers in the
curly brackets stand for the labels, according to
Fig.~\ref{blocks}, of the minimizing $T$-plaquette
configurations, and the $T$-plaquette configurations that can be
obtained from the named ones, by applying symmetries of $H_{0}$,
are omitted. Consequently, the sets of ground-state
configurations in the five open domains are given by:
${\mathcal{S}}_{-}=\{S_{-}\}$,
${\mathcal{S}}_{cb}=\{S_{cb}^{e},S_{cb}^{o}\}$,
${\mathcal{S}}_{1}=\{S_{1}^{j}\}_{j=1}^{j=10}$,
${\mathcal{S}}_{2}=\{S_{2}^{j}\}_{j=1}^{j=8}$, and
${\mathcal{S}}_{3}=\{S_{3}^{j}\}_{j=1}^{j=6}$, where  the
representative configurations of ${\mathcal{S}}_{cb}$,
${\mathcal{S}}_{1}$, ${\mathcal{S}}_{2}$, and
${\mathcal{S}}_{3}$, are displayed in Fig.~\ref{configurations}.
The coefficients $\alpha_{i}$ are given in the Table~\ref{tab1}
included in Appendix.

For $\varepsilon=0$ the obtained phase diagram coincides with the
phase diagram of the spinless Falicov--Kimball model
\cite{GJL,GMMU}. If  $\varepsilon<12$, when $\delta$ is increased,
it crosses some ``critical'' values, $\delta_{cr}(\varepsilon)$,
that limit the $\delta$-extent of the domains ${\mathcal{S}}_{-}$,
${\mathcal{S}}_{1}$, ${\mathcal{S}}_{2}$, ${\mathcal{S}}_{3}$, and
${\mathcal{S}}_{cb}$. In particular, $\delta_{cr}(0)=-1/2$, $2$,
$6$, $21/2$, respectively. Similarly, as $\varepsilon$ is
increased, it passes through three ``critical'' values
$\varepsilon_{cr}=4$, $8$, $12$, that limit the
$\varepsilon$-extent of the domains ${\mathcal{S}}_{2}$,
${\mathcal{S}}_{1}$, ${\mathcal{S}}_{3}$, respectively. Thus,
there is no ${\mathcal{S}}_{2}$ ground-state configurations for
$\varepsilon>4$, etc. Above $\varepsilon=12$ the phase diagram is
independent of $\varepsilon$.

At the boundaries of the domains the situation is more involved.
Let ${\mathcal{S}}$, ${\mathcal{S^{\prime}}}$ be two domains of
the considered phase diagram. The set of the minimizing
$T$-plaquette configurations at the boundary between
${\mathcal{S}}$ and ${\mathcal{S^{\prime}}}$ consists always of
${{\mathcal{S}}_{|T}} \cup {{\mathcal{S^{\prime}}}_{|T}}$, but it
may contain also some additional $T$-plaquette configurations of
minimal energy. If it is the case, a great many of ground-state
configurations exists at the boundary. Except the boundary
between ${\mathcal{S}}_{-}$ and ${\mathcal{S}}_{cb}$, the number
of ground-state configurations that can be built out of the
minimizing $T$-plaquette configurations grows indefinitely with
the size of the lattice.

Specifically, at the boundary between ${\mathcal{S}}_{-}$ and
${\mathcal{S}}_{cb}$ the set of the minimizing $T$-plaquette
configurations is ${\mathcal{S}}_{-|T} \cup {\mathcal{S}}_{cb|T}$.
Clearly, only the three periodic configurations, $S_{-}$,
$S_{cb}^{e}$, and $S_{cb}^{o}$, can be built out of them.

Then, at the boundary between ${\mathcal{S}}_{-}$ and
${\mathcal{S}}_{3}$ the set of the minimizing $T$-plaquette
configurations is ${\mathcal{S}}_{-|T} \cup {\mathcal{S}}_{3|T}
\cup \{ 4,7,9 \}$. A simple reasoning shows that in any
ground-state configuration all the lattice lines with one of the
slopes $1$ or $-1$ are ordered ferromagnetically. On any
horizontal line of sites, every two consecutive sites with
pseudo-spins $1$ are separated by at least two sites with
pseudo-spins $-1$. That is, the ground-state configurations at
the boundary ``interpolate'' between the configurations
${\mathcal{S}}_{3}$ and ${\mathcal{S}}_{-}$. The number of such
configurations grows as $\exp({\mathrm{const}} \sqrt{|\Lambda|})$.
The last segment of the boundary of ${\mathcal{S}}_{-}$ is the
boundary between ${\mathcal{S}}_{-}$ and ${\mathcal{S}}_{1}$. The
set of the minimizing $T$-plaquette configurations is
${\mathcal{S}}_{-|T} \cup {\mathcal{S}}_{1|T} \cup \{ 3,4,9 \}$.
Therefore, flipping in ${\mathcal{S}}_{1}$ any set of
pseudo-spins $1$, we obtain a ground-state configuration. Their
number grows as $\exp( {\mathrm{const}} |\Lambda|)$.

The situation at the boundary between ${\mathcal{S}}_{1}$ and
${\mathcal{S}}_{3}$ is more intricate. The set of the minimizing
$T$-plaquette configurations is ${\mathcal{S}}_{1|T} \cup
{\mathcal{S}}_{3|T} \cup \{ 6 \}$. In any ground-state
configuration the lattice lines with one of the slopes $2$,
$1/2$, $-2$, or $-1/2$ are ordered ferromagnetically. Along such
a line, it is possible to build a stratum, of any width, that
consists of elementary cells of ${\mathcal{S}}_{1}$ kind
($\sqrt{5} \times \sqrt{5}$ squares) (see
Fig.~\ref{configurations}), then a stratum that consits of
elementary cells of ${\mathcal{S}}_{3}$ kind (diamonds), and so
on. The number of such periodic configurations grows as
$\exp({\mathrm{const}} \sqrt{|\Lambda|})$. Moreover, it is
possible to build quasi-periodic tilings composed of square
elementary cells and diamonds \cite{Watson}.

The three boundaries that remain to be described, have already
been discussed in the literature. At the boundary between
${\mathcal{S}}_{3}$ and ${\mathcal{S}}_{cb}$, the set of the
minimizing $T$-plaquette configurations is ${\mathcal{S}}_{3|T}
\cup {\mathcal{S}}_{cb|T} \cup \{ 19 \}$. In any ground-state
configuration the lattice lines with slope $1$ or $-1$ are ordered
ferromagnetically. On every horizontal and every vertical line of
sites, any two consecutive sites with pseudo-spins $1$ are
separated by either one or two sites with pseudo-spins equal to
$-1$ \cite{Kennedy1}. Thus, the number of such configurations
grows as $\exp({\mathrm{const}} \sqrt{|\Lambda|})$.

Next, at the boundary between ${\mathcal{S}}_{2}$ and
${\mathcal{S}}_{3}$ the set of the minimizing $T$-plaquette
configurations is ${\mathcal{S}}_{2|T} \cup {\mathcal{S}}_{3|T}
\cup \{ 6 \}$. In any ground-state configuration the lattice lines
with one of the slopes $2$, $1/2$, $-2$, or $-1/2$ are ordered
ferromagnetically. If the slope is $\pm 2$, on every vertical
line of sites, any two consecutive sites with pseudo-spins $1$
are separated by either two or three sites with pseudo-spins
$-1$. If the slope is $\pm 1/2$, on every horizontal line any two
consecutive sites with pseudo-spins $1$ are separated by either
two or three sites with pseudo-spins $-1$ \cite{Kennedy1}. The
number of such configurations grows as $\exp( {\mathrm{const}}
\sqrt{|\Lambda|})$.

Finally at the boundary between ${\mathcal{S}}_{1}$ and
${\mathcal{S}}_{2}$, the set of the minimizing $T$-plaquette
configurations is ${\mathcal{S}}_{1|T} \cup {\mathcal{S}}_{2|T}$.
Here the situation is analogous to that of the boundary between
${\mathcal{S}}_{1}$ and ${\mathcal{S}}_{3}$ \cite{Watson}.

\subsection{Construction of the phase diagram up to fourth order.
The case of hopping hard-core bosons}

\begin{figure}
\centering \includegraphics[width=\textwidth]{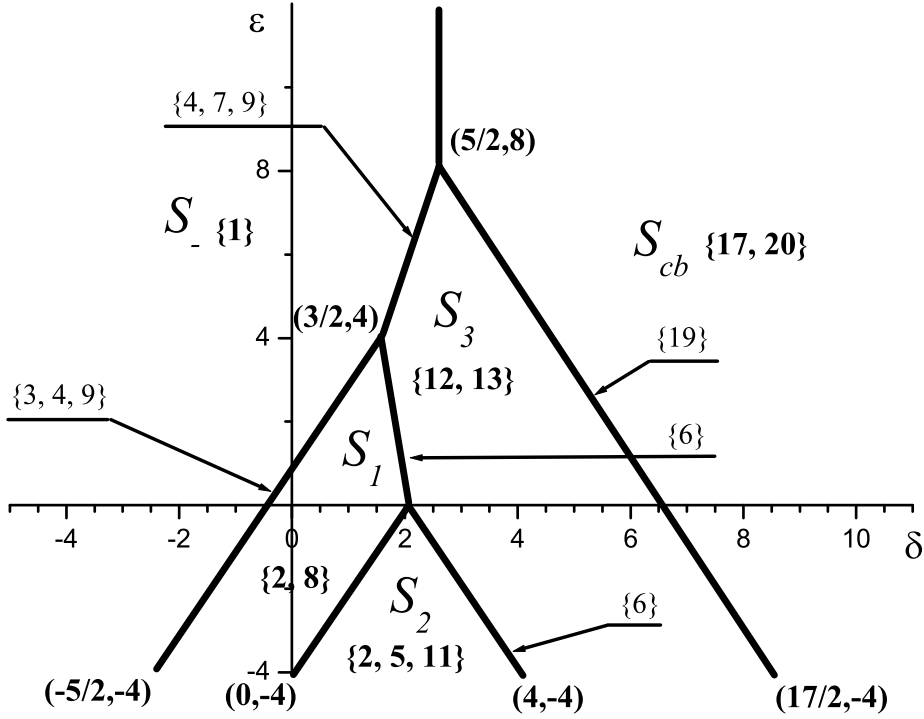}
\caption{Ground-state phase diagram of
${\tilde{H}}^{(4)}_{{\mathrm{eff}}}$. See description of
Fig.~\ref{t4}.} \label{t4b}
\end{figure}

In the case of hard-core bosons, the creation and annihilation
operators at site $x$: $c^{+}_{x}$, $c_{x}$, respectively, satisfy
the anticommutation relations, as fermions, but in contrast to
fermions they commute at different sites. Consequently, with the
system of hard-core bosons, described by the Hamiltonian
(\ref{H0t/U}), we cannot associate a one-particle Hamiltonian and
derive the small $t$ effective interaction  as in the case of
fermions. The effective interaction up to fourth order has been
derived in \cite{GMMU} by means of a closed-loop expansion, and
reads:
\begin{eqnarray}
\label{hbexpansion} {\tilde{E}}_{S} \left( \mu_e,\mu_i \right)
&=& -\frac{1}{2} \left( \mu_{i}-\mu_{e} \right) \sum\limits_{x}
s_{x} -\frac{1}{2} \left( \mu_{i}+\mu_{e}+1 \right) |\Lambda| +
\nonumber \\
&&\left[ \frac{t^{2}}{4}- \frac{5t^{4}}{16} \right]
\sum\limits_{\langle x,y \rangle_{1}}  s_{x}s_{y} + \left[
\frac{5t^{4}}{16}- \frac{{\tilde{\varepsilon}}}{16} \right]
\sum\limits_{\langle x,y \rangle_{2}}  s_{x}s_{y} +
\nonumber \\
&&\frac{t^{4}}{8} \sum\limits_{\langle x,y \rangle_{3}} s_{x}s_{y}
- \frac{t^{4}}{16} \sum\limits_{P} \left(5+s_{P}\right) +
{\tilde{R}}^{(4)}.
\end{eqnarray}
The above expansion is convergent absolutely if $t<1/16$ and
$\left| \mu_e \right|< 1 - 16t$. The latter condition implies that
$N_{e} + N_{i} = |\Lambda|$. Setting in (\ref{hbexpansion})
$\mu_{e}=0$, $\mu_{i} =-2t^{2}+ \delta t^{4}$, and dropping the
remainder ${\tilde{R}}^{(4)}$, we get the effective interaction
up to fourth order, ${\tilde{E}}_{S}^{(4)} \left( 0,\delta
\right)$, and the effective Hamiltonian,
${\tilde{H}}^{(4)}_{{\mathrm{eff}}}$, defined as in (\ref{E4}).
Then, repeating all the steps of the construction of the phase
diagram up to fourth order, carried out for fermions in the
previous section, leads to the phase diagram displayed in
Fig.~\ref{t4b}.
The diagram in Fig.~\ref{t4b} is described in the same way as the
diagram in Fig.~\ref{t4}. If the phase diagram due to
${\tilde{E}}_{S}^{(4)} \left( 0,\delta \right)$ is limited to
positive $\varepsilon$, then it differs significantly from its
counterpart for fermions, the ${\mathcal{S}}_{2}$-domain is
missing. But the analysis extended to negative values of
$\varepsilon$ reveals the ${\mathcal{S}}_{2}$-domain, and shows
that the phase diagram of hard-core bosons, limited to
$\varepsilon \geqslant -4$, after translating by the vector
$(2,4)$ becomes identical with the phase diagram of fermions,
limited to $\varepsilon \geqslant 0$.

\section{Discussion of the phase diagram and summary}

The ground-state phases of the whole electron-ion system under
consideration can be distinguished by their ground-state
arrangements of ions. Thus, we can speak of
${\mathcal{S}}_{cb}$-phase, where the arrangement of the ions is
given by the chessboard configurations , etc. Let us ignore for
the moment the remainder $R^{(4)}$ in the expansion
(\ref{expansion}). Then, it follows from the grand-canonical
phase diagram due to the effective interaction truncated at
fourth order that if $\rho_{i}=\rho_{e}=1/2$, then the
ground-state phase is the ${\mathcal{S}}_{cb}$-phase. More
generally, consider a sequence of rational ion densities $\rho_i=$
$0$, $1/5$, $1/4$, $1/3$, $1/2$ (and the corresponding electron
densities $\rho_{e}=1-\rho_{i}$). Then, for some densities of this
sequence, depending on $\varepsilon$, the ground-state phases of
the system are characterized by crystalline, \ie{} exibiting
positional long-range order, arrangements of the ions, given by
the ion configurations ${\mathcal{S}}_{-}$, ${\mathcal{S}}_{1}$,
${\mathcal{S}}_{2}$, ${\mathcal{S}}_{3}$ and ${\mathcal{S}}_{cb}$,
of the corresponding density. For $\varepsilon
> 12$, we can have only two crystalline phases: ${\mathcal{S}}_{-}$-phase and
${\mathcal{S}}_{cb}$-phase. At the half-line, given by
$\delta=9/2$ and $\varepsilon > 12$, these two phases and only
these two ones coexist, \ie{} any $S$ that is not in
${\mathcal{S}}_{-} \cup {\mathcal{S}}_{cb}$ has higher energy.
Since,
\begin{eqnarray}
\label{E4,9/2} E_{S}^{(4)}\left( 0,\frac{9}{2} \right) + t^{4}
\left( \frac{\varepsilon}{8} + \frac{11}{64}
\right)|\Lambda| = \nonumber\\
\frac{t^{4}}{2}
H^{(4)}_{{\mathrm{eff}}}+t^{2}\sum\limits_{\substack{\langle x,y
\rangle_{1} \\
s_{x}=s_{y}=1}} 1 +t^{4} \left( \frac{\varepsilon}{8} +
\frac{11}{64}
\right)|\Lambda| = \nonumber \\
\frac{t^{4}}{2} \sum\limits_{T} \left(
H_{T}^{(4)}+K_{T}^{(4)}-E^{(-,cb)}_{T} \right)
+t^{2} \sum\limits_{\substack{\langle x,y \rangle_{1} \\
s_{x}=s_{y}=1}} 1,
\end{eqnarray}
where $H_{{\mathrm{eff}}}^{(4)}$, $H_{T}^{(4)}$, and
$K_{T}^{(4)}$ are evaluated at $\delta=9/2$, and $E^{(-,cb)}_{T}$
stands for the common value of $H_{T}^{(4)} + K_{T}^{(4)}$
obtained for $S \in {\mathcal{S}}_{-} \cup {\mathcal{S}}_{cb}$
and $\delta=9/2$, the coexistence of ${\mathcal{S}}_{-}$-phase and
${\mathcal{S}}_{cb}$-phase follows from the following inequality:
\begin{eqnarray}
\label{E4bound} \frac{t^{4}}{64}(\varepsilon - 12)
\left| {\mathcal{T}}_{-,cb}(S) \right| \leqslant \nonumber \\
\frac{t^{4}}{2} \sum\limits_{T} \left( H_{T}^{(4)}+K_{T}^{(4)} -
E^{(-,cb)}_{T} \right)
+t^{2} \sum\limits_{\substack{\langle x,y \rangle_{1} \\
s_{x}=s_{y}=1}} 1 \leqslant \nonumber \\
{\mathcal{C}}^{(4)} t^{2} \left| {\mathcal{T}}_{-,cb}(S) \right|,
\end{eqnarray}
for any configuration $S$ and $\varepsilon > 12$. In
(\ref{E4bound}), $H_{T}^{(4)}$ and $K_{T}^{(4)}$ are also
evaluated at $\delta=9/2$, ${\mathcal{T}}_{-,cb}(S)$ stands for
the set of the $T$-plaquettes such that $S_{|T}$ does not coincide
with $T$-plaquette configurations labelled by $1$, $17$, $20$ (we
call ${\mathcal{T}}_{-,cb}(S)$  the set of {\em $T$-plaquette
excitations\/} relative to the configurations in
${\mathcal{S}}_{-} \cup {\mathcal{S}}_{cb})$,
$|{\mathcal{T}}_{-,cb}(S)|$ is the cardinality of
${\mathcal{T}}_{-,cb}(S)$, and ${\mathcal{C}}^{(4)}$ is a
constant (\ie{} ${\mathcal{C}}^{(4)}$ is independent of $S$, $t$,
$\varepsilon$, and $\Lambda$). The inequality (\ref{E4bound}),
together with the variational argument constructed in
\cite{Kennedy2}, can be used to prove that for any density $0<
\rho_{i} < 1/2$, the ground-state phase is a mixture (a state of
phase separation) of the ${\mathcal{S}}_{-}$-phase and
${\mathcal{S}}_{cb}$-phase. Specifically, the variational
argument shows that for $\varepsilon > 12$ there is a constant
$c^{(4)}$ such that in any ground-state the number $\left|
{\mathcal{T}}_{-,cb}(S) \right|$ of $T$-plaquette excitations is
bounded from above as follows:
\begin{equation}
\label{T4bound} \left| {\mathcal{T}}_{-,cb}(S) \right| \leqslant
\frac{c^{(4)} t^{-2}}{\varepsilon-12} \sqrt{|\Lambda|}.
\end{equation}
This means that, for $0< \rho_{i} < 1/2$ and sufficiently large
lattice, any ground-state configuration of ions consists of
connected regions whose ion configurations are restrictions to
these regions of one of the configurations $S_{-}$, $S_{cb}^{e}$,
$S_{cb}^{o}$. These connected regions are separated by domain
walls that involve negligible, \ie{} $O(|\Lambda|^{-1/2})$ (for
large lattice $\Lambda$), fraction of sites.

Now, let us fix the ion density, $\rho_{i}=1/5$ or $1/4$ or $1/3$,
and a sufficiently small $t$. If we choose an initial value of
${\tilde{\varepsilon}}$ so that
$\varepsilon={\tilde{\varepsilon}}/t^{4}$ is greater than $12$
(\ie{} initially the system is in the state of phase-separation),
then by decreasing ${\tilde{\varepsilon}}$ sufficiently the
system is driven into a crystalline state. But obviously, the
same result can be arrived at, if we fix the value of
${\tilde{\varepsilon}}$ and an (sufficiently small) initial value
of $t_{0}$ so that
$\varepsilon={\tilde{\varepsilon}}/t_{0}^{4}=13$, for instance,
and then increase $t$. If not the fact that we have ignored the
remainder $R^{(4)}$ in the effective interaction, this would be
the announced in the Introduction statement. Thus, it remains to
be proven that the remainder $R^{(4)}$ does not destroy the
picture obtained with the truncated effective interaction
$E_{S}^{(4)}$.

First, we shall demonstrate that if we take into account $R^{(4)}$,
then there is a (sufficiently small) constant $t_{0}$ such that for
$t<t_{0}$ the sets of ground-state configurations of ions for
densities $\rho_{i}=1/5$, $1/4$, $1/3$ coincide with the sets
${\mathcal{S}}_{1}$, ${\mathcal{S}}_{2}$, and ${\mathcal{S}}_{3}$,
respectively. Specifically, we shall show that for $t<t_{0}$ there
are nonempty two-dimensional open domains
${\mathcal{S}}_{1}^{\infty}$, ${\mathcal{S}}_{2}^{\infty}$, and
${\mathcal{S}}_{3}^{\infty}$ that are contained in the domains
${\mathcal{S}}_{1}$, ${\mathcal{S}}_{2}$, and ${\mathcal{S}}_{3}$,
respectively, and such that in ${\mathcal{S}}_{j}^{\infty}$,
$j=1,2,3$, the set of ground-state configurations is
${\mathcal{S}}_{j}$. To achieve this we shall construct a lower
bound for the energy difference $E_{S} \left( 0,-2t^{2} + \delta
t^{4} \right) - E_{S_{j}} \left( 0,-2t^{2} + \delta t^{4} \right)$,
with $S_{j} \in {\mathcal{S}}_{j}$. Clearly, we can restrict the set
of all configurations $S$ to those that do not contain any pairs
$\langle x,y \rangle_{1}$ with $s_{x}=s_{y}=1$. With this
restriction,
\begin{eqnarray}
\label{Ediff} E_{S} \left( 0,-2t^{2} + \delta t^{4} \right)-
E_{S_{j}} \left( 0,-2t^{2} +
\delta t^{4} \right) = \nonumber \\
\frac{t^{4}}{2}\sum\limits_{T \in {\mathcal{T}}_{j}(S)} \left[
\left. \left( H_{T}^{(4)}+K_{T}^{(4)} \right) \right|_{S} - \left.
\left( H_{T}^{(4)}+K_{T}^{(4)} \right) \right|_{S_{j}} \right] +
\left. R^{(4)} \right|_{S}- \left. R^{(4)} \right|_{S_{j}}.
\end{eqnarray}
Clearly,
\begin{eqnarray}
\label{E4diff} \sum\limits_{T \in {\mathcal{T}}_{j}(S)} \left[
\left. \left( H_{T}^{(4)}+K_{T}^{(4)} \right) \right|_{S} - \left.
\left( H_{T}^{(4)}+K_{T}^{(4)} \right) \right|_{S_{j}} \right]
\geqslant  \tau_{j} \left| {\mathcal{T}}_{j}(S) \right|
\end{eqnarray}
where ${\mathcal{T}}_{j}(S)$ is the set of $T$-plaquette
excitations relative to the ground-state configurations from
${\mathcal{S}}_{j}$ and
\begin{equation}
\label{tau} \tau_{j} = \min \left\{ \left. \left(
H_{T}^{(4)}+K_{T}^{(4)} \right) \right|_{S} - \left. \left(
H_{T}^{(4)}+K_{T}^{(4)} \right) \right|_{S_{j}}: S \notin
{\mathcal{S}}_{j} \right\}
\end{equation}
is a function of $(\delta,\varepsilon)$ on the domain
${\mathcal{S}}_{j}$. Following  \cite{GMMU} we have also the
upper bound:
\begin{equation}
\label{R4diff} \left| \left. R^{(4)} \right|_{S}- \left. R^{(4)}
\right|_{S_{j}} \right| \leqslant t^{6} r_{j} \left|
{\mathcal{T}}_{j}(S) \right|
\end{equation}
for some constant $r_{j}$. From
(\ref{Ediff},\ref{E4diff},\ref{R4diff}) we obtain
\begin{equation}
\label{Elb} E_{S} \left( 0,-2t^{2} + \delta t^{4} \right)- E_{S_{j}}
\left( 0,-2t^{2} + \delta t^{4} \right) \geqslant t^{4} \left(
\tau_{j}- r_{j} t^{2} \right) \left| {\mathcal{T}}_{j}(S) \right| .
\end{equation}
Note that ${\mathcal{S}}_{j}$ is an open convex domain in the
$(\delta,\varepsilon)$-plane. For $(\delta,\varepsilon) \in
{\mathcal{S}}_{j}$, the function $(\delta,\varepsilon)
\rightarrow \tau_{j}(\delta,\varepsilon)$, being the minimum of a
finite set of positive linear functions, is concave, piecewise
linear and positive. It vanishes only at the boundary of
${\mathcal{S}}_{j}$. Therefore, there is a (sufficiently small)
$t_{0}$  such that for $t<t_{0}$ the value of $r_{j} t^{2}$ is
less than the maximum of $\tau_{j}$ over ${\mathcal{S}}_{j}$.
Then, the level set $\left\{ (\delta,\varepsilon):
\tau_{j}(\delta,\varepsilon) > r_{j} t^{2} \right\}$ is a nonempty
convex open set, contained in domain ${\mathcal{S}}_{j}$. At this
level set the ground-state configurations coincide with the
configurations from ${\mathcal{S}}_{j}$.

Second, we shall show that for a sufficiently small $t$ the
inequality (\ref{T4bound}) is modified by $R^{(4)}$ only slightly,
so that the conclusions drawn from (\ref{T4bound}) remain valid.
The remainder $R^{(4)}$ can be estimated in the following way
\cite{Kennedy2}:
\begin{equation}
\label{R4bound} \left| \left. R^{(4)} \right|_{S} - \left. R^{(4)}
\right|_{S_{-}} - \left. R^{(4)} \right|_{S_{cb}} \right|
\leqslant t^{6} r_{-,cb} \left| {\mathcal{T}}_{-,cb}(S) \right|,
\end{equation}
for some constant $r_{-,cb}$. Therefore, taking into account the
remainder $R^{(4)}$, we replace the inequality (\ref{E4bound}) by
\begin{eqnarray}
\label{Ebound} \frac{t^{4}}{64}[\varepsilon - (12 + t^{2}
r_{-,cb})] |{\mathcal{T}}_{-,cb}(S)|
& \leqslant &  \nonumber \\
\frac{t^{4}}{2} \sum\limits_{T} \left( H_{T}^{(4)}+K_{T}^{(4)} -
E^{(-,cb)}_{T} \right)
+ & & \nonumber \\
t^{2} \sum\limits_{\substack{\langle x,y \rangle_{1} \\
s_{x}=s_{y}=1}} 1 + R^{(4)} - \left. R^{(4)} \right|_{S_{-}} -
\left. R^{(4)} \right|_{S_{cb}} & \leqslant & {\mathcal{C}} t^{2}
|{\mathcal{T}}_{-,cb}(S)|,
\end{eqnarray}
for some constant ${\mathcal{C}}$. As a result, the variational
argument of \cite{Kennedy2} implies that there is a constant $c$
such that the following counterpart of (\ref{T4bound}) holds:
\begin{equation}
\label{Tbound} |{\mathcal{T}}_{-,cb}(S)| \leqslant \frac{c
t^{-2}}{\varepsilon-(12 + 64 r_{-,cb} t^{2} )} \sqrt{|\Lambda|}.
\end{equation}
Consequently, for any density $0< \rho_{i} < 1/2$ and \nnn{}
interaction $\varepsilon > 12 + 64 r_{-,cb} t^{2}$ the
ground-state phase is a mixture of the ${\mathcal{S}}_-$-phase and
${\mathcal{S}}_{cb}$-phase.

Summing up, let the ion density be one of $\rho_{i}=1/5$, $1/4$,
and $1/3$, and let the corresponding electron density be
$\rho_{e}=1- \rho_{i}$. We restrict our considerations to the case
$\rho_{i}<1/2$, since the situation for $\rho_{i}>1/2$ can be
obtained by means of symmetries (see subsection 2.1). Let the
hopping intensity $t$ be sufficiently small. Then, we have proved
that the considered system can be driven from the state of phase
separation (the mixture of ${\mathcal{S}}_{-}$- and
${\mathcal{S}}_{cb}$-phases) to a crystalline state with a
long-range order, which is the ${\mathcal{S}}_{j}$-phase whose
particle densities are the chosen ones, by increasing quantum
fluctuations (measured by $t$) due to the hopping electrons.

Finally, it is easy to see that if the considered model is
augmented by a \nn{} repulsion between the ions, then the above
analysis of the phase diagram can be carried out without any
essential modifications. Thus, we arrive at the same conclusions
concerning the transition from a state of phase separation to a
state with a long-range order due to increasing hopping.

 A similar discussion can be carried out in the case
of hard-core bosons, with one exception. Our analysis shows that
the transition from a state of phase separation to a state with a
long-range order due to an increasing hopping occurs for densities
$\rho_{i}=1/5$, $1/3$.

\ack It is a pleasure to thank Romuald Lema{\'{n}}ski for
discussions. V.D. is grateful to the University of Wroc{\l}aw,
and especially to the Institute of Theoretical Physics for
financial support.

\section*{Appendix}
Here we provide the tables of the coefficients $\alpha_i$, $i=1,
\ldots, 5$, of the zero-potentials in the case of hopping fermions
and hopping hard-core bosons.
\begin{table}[ht]
\scriptsize
\begin{center}
\caption{Zero-potential coefficients $\alpha_{i}$ for fermions
($\varepsilon \geqslant 0$)} \label{tab1}
\begin{tabular}{|c|c|c|c|c|c|}
\hline & $\alpha_{1}$ & $\alpha_{2}$ & $\alpha_{3}$ &
$\alpha_{4}$ & $\alpha_{5}$
\\
\hline $S_{-}$, $\varepsilon \leqslant 8$ & $0$ &
$-\frac{\delta}{8}$ & $\frac{\delta}{32}+\frac{1}{64}$ & $0$ &
$-\frac{1}{48}$
\\
\hline $S_{-}$, $\varepsilon \geqslant 8$ & $0$ &
$-\frac{\delta}{8}$ &
$\frac{\delta}{32}+\frac{\varepsilon}{128}-\frac{3}{64}$ & $0$ &
$-\frac{1}{48}$
\\
\hline $S_{1}$ & $-\frac{3 \delta}{40}+ \frac{\varepsilon}{160}
-\frac{3}{80}$ & $-\frac{\delta}{10}- \frac{\varepsilon}{80}
+\frac{1}{80}$ & $0$ & $-\frac{\delta}{40}+ \frac{\varepsilon}{80}
-\frac{1}{80}$ & $-\frac{1}{48}$
\\
\hline $S_{2}$ & $-\frac{7 \delta}{128}- \frac{\varepsilon}{256}-
\frac{5}{64}$ & $-\frac{9 \delta}{64}+ \frac{\varepsilon}{128}+
\frac{3}{32}$ & $0$ &
$\frac{\delta}{128}-\frac{\varepsilon}{256}- \frac{5}{64}$ &
$\frac{\delta}{128}- \frac{ \varepsilon}{256} - \frac{7}{192}$
\\
\hline $S_{3}$, $\varepsilon \leqslant 4$ & $-\frac{11
\delta}{144}- \frac{\varepsilon}{144}+ \frac{5}{96}$ &
$-\frac{\delta}{8}$ & $0$ & $\frac{\delta}{144}+
\frac{\varepsilon}{288}- \frac{7}{96}$ & $-\frac{\delta}{144}-
\frac{\varepsilon}{288}+ \frac{5}{96}$
\\
\hline $S_{3}$, $4 \leqslant \varepsilon \leqslant 8$ &
$-\frac{\delta}{12}-\frac{\varepsilon}{384}+\frac{1}{16}$ &
$-\frac{\delta}{8}$ & $0$ &
$\frac{\varepsilon}{128}-\frac{1}{16}$ &
-$\frac{\varepsilon}{128}+\frac{1}{24}$
\\
\hline $S_{3}$, $\varepsilon \geqslant 8$ &
$-\frac{\delta}{12}-\frac{\varepsilon}{96}+\frac{1}{8}$ &
$-\frac{\delta}{8}$ & $0$ & $0$ & $-\frac{1}{48}$
\\
\hline $S_{cb}$ & $0$ & $-\frac{\delta}{8}$ & $\frac{185 \delta
}{5856}+ \frac{\varepsilon}{5856}+ \frac{169}{3904}$ & $0$ &
$-\frac{1}{48}$
\\ \hline
\end{tabular}
\end{center}
\end{table}
\begin{table}[ht]
\scriptsize
\begin{center}
\caption{Zero-potential coefficients $\alpha_{i}$ for hard-core
bosons ($\varepsilon \geqslant -4$)} \label{tab2}
\begin{tabular}{|c|c|c|c|c|c|}
\hline & $\alpha_{1}$ & $\alpha_{2}$ & $\alpha_{3}$ &
$\alpha_{4}$ & $\alpha_{5}$
\\
\hline $S_{-}$, $\varepsilon \leqslant 4$ & $0$ &
$-\frac{\delta}{8}$ & $\frac{\delta}{32}- \frac{1}{192}$ & $0$ &
$-\frac{1}{48}$
\\
\hline $S_{-}$, $\varepsilon \geqslant 4$ & $0$ &
$-\frac{\delta}{8}$ & $\frac{\delta}{32}+\frac{\varepsilon}{128}-
\frac{7}{192}$ & $0$ & $-\frac{1}{48}$
\\
\hline $S_{1}$ & $-\frac{3 \delta}{40}+ \frac{\varepsilon}{160}
+\frac{1}{240}$ & $-\frac{\delta}{10}- \frac{\varepsilon}{80}
+\frac{1}{80}$ & $0$ & $-\frac{\delta}{40}+ \frac{\varepsilon}{80}
-\frac{1}{80}$ & $-\frac{1}{48}$
\\
\hline $S_{2}$ & $-\frac{7 \delta}{128}- \frac{\varepsilon}{256}-
\frac{7}{192}$ & $-\frac{9 \delta}{64}+ \frac{\varepsilon}{128}+
\frac{3}{32}$ & $0$ &
$\frac{\delta}{128}-\frac{\varepsilon}{256}- \frac{5}{64}$ &
$\frac{\delta}{128}- \frac{ \varepsilon}{256} - \frac{7}{192}$
\\
\hline $S_{3}$, $\varepsilon \leqslant 0$ & $-\frac{11
\delta}{144}- \frac{\varepsilon}{144}+ \frac{11}{288}$ &
$-\frac{\delta}{8}$ & $0$ & $\frac{\delta}{144}+
\frac{\varepsilon}{288}- \frac{13}{288}$ & $-\frac{\delta}{144}-
\frac{\varepsilon}{288}+ \frac{7}{288}$
\\
\hline $S_{3}$, $0 \leqslant \varepsilon \leqslant 4$ &
$-\frac{\delta}{12}-\frac{\varepsilon}{384}+\frac{5}{96}$ &
$-\frac{\delta}{8}$ & $0$ &
$\frac{\varepsilon}{128}-\frac{1}{32}$ &
-$\frac{\varepsilon}{128}+\frac{1}{96}$
\\
\hline $S_{3}$, $\varepsilon \geqslant 4$ &
$-\frac{\delta}{12}-\frac{\varepsilon}{96}+\frac{1}{12}$ &
$-\frac{\delta}{8}$ & $0$ & $0$ & $-\frac{1}{48}$
\\
\hline $S_{cb}$ & $0$ & $-\frac{\delta}{8}$ & $\frac{185 \delta
}{5856}+ \frac{\varepsilon}{5856}+ \frac{93}{3904}$ & $0$ &
$-\frac{1}{48}$
\\ \hline
\end{tabular}
\end{center}
\end{table}

\end{document}